\documentclass[pdflatex,sn-mathphys-num]{sn-jnl}
\usepackage{amsmath,amssymb,amstext,braket, graphicx,color,xfrac,accents,tabulary,booktabs,xcolor,subcaption,cases}
\usepackage{multirow}%
\usepackage{amsthm}%
\usepackage{mathrsfs}%
\usepackage[title]{appendix}%
\usepackage{xcolor}%
\usepackage{textcomp}%
\usepackage{manyfoot}%
\usepackage{booktabs}%
\usepackage{algorithm}%
\usepackage{algorithmicx}%
\usepackage{algpseudocode}%
\usepackage{listings}%

\usepackage[T1]{fontenc}
\usepackage[utf8]{inputenc}

\newtheorem{thm}{Theorem}[section]

\newtheorem{example}[thm]{Example}

\newcommand{\hs}{\mathcal{H}}

\newcommand{\hst}{\mathcal{H}_{t}}

\newcommand{\be}{\begin{equation}}
\newcommand{\ee}{\end{equation}}

\newcommand{\ham}{\hat{H}}
\newcommand{\lcb}{\left[}
\newcommand{\rcb}{\right]}

\newcommand{ \toptr }{\hat{t}}
\newcommand{ \pt }{\hat{p}_t}

\newcommand{\eye}{\hat{\mathbb{I}}}
\newcommand{ \J }{\hat{\mathbb{J}}}

\newcommand{\kett}[1]{\left| #1 \right\rangle\bigr\rangle}
\newcommand{\braa}[1]{ \left\langle\bigl\langle  #1 \right|}

\def\({\left(}
\def\){\right)}
\def\[{\left[}
\def\]{\right]}

\definecolor{purple}{rgb}{0.5,0,0.5}

\begin{document}

\title[Emergence of Gravitational Potential and Time Dilation from Non-interacting Systems Coupled to a Global Quantum Clock]{Emergence of Gravitational Potential and Time Dilation from Non-interacting Systems Coupled to a Global Quantum Clock}


\author*[1,2]{\fnm{Ashmeet} \sur{Singh}}\email{ashmeet@physics.iitd.ac.in}

\author[3,4]{\fnm{Oliver} \sur{Friedrich}}\email{oliver.friedrich@physik.uni-muenchen.de}
\equalcont{These authors contributed equally to this work.}

\affil*[1]{\orgdiv{Department of Physics}, \orgname{Indian Institute of Technology Delhi}, \orgaddress{\street{Hauz Khaz}, \city{New Delhi}, \postcode{110016}, \country{India}}}
\affil*[2]{\orgdiv{Physics Department}, \orgname{Whitman College}, \orgaddress{\street{345 Boyer Ave}, \city{Walla Walla}, \postcode{99362}, \state{WA}, \country{USA}}}

\affil[3]{\orgdiv{Faculty for Physics, University Observatory}, \orgname{Ludwig-Maximilians Universität München}, \orgaddress{\street{Scheinerstr.\ 1}, \city{Munich}, \postcode{81677}, \country{Germany}}}

\affil[4]{\orgdiv{Excellence Cluster ORIGINS},\orgaddress{\street{Boltzmannstr.\ 2}, \city{Garching}, \postcode{85748},\country{Germany}}}


\abstract{
We study gravitational back-reaction within the Page-Wootters formulation of quantum mechanics by treating time as a quantum degree of freedom. Our model introduces a distinction between global ``coordinate time,'' represented as a relational quantum observable, and ``proper time,'' measured by internal quantum degrees of freedom of physical systems. By coupling mass-energy with coordinate time through a Wheeler-DeWitt-like constraint, we demonstrate the natural emergence of gravitational time dilation. In the presence of a massive object this agrees with time dilation in a Schwarzchild metric at leading order if the interaction strength is taken to be representative of the gravitational coupling $G$. Additionally, when two particles independently couple to the time coordinate, a Newtonian gravitational interaction arises in the low-energy limit, showing how gravitational potential can emerge from non-interacting quantum systems. Our approach also reveals renormalization features, potentially softening high-energy divergences and suggesting that particles in superposition might introduce quantum corrections to gravitational time dilation. }




\maketitle

\maketitle


\section{Introduction} \label{sec:intro}

A defining feature of a complete quantum theory of gravity will be the ability to treat space and time as quantum degrees of freedom. This will include, among other things, allowing for superpositions of classical spacetime geometries and a model for the quantum degrees of freedom which make up spacetime. Treating both time and space on a quantum footing is challenging, but it allows for the possibility of quantum mechanical interactions between matter and spacetime degrees of freedom, from which gravity may emerge. In such approaches \cite{Carroll:2018rhc, giddings2018quantum,giddings2019quantum} to quantum gravity, one begins with minimal quantum mechanical elements in Hilbert space and from it, derive higher-level structures such as space, locality, matter, and eventually, gravity. Approaches to quantum gravity are often plagued by the ``problem of time" \cite{Isham:1992ms, anderson_problem_time, kuchavr1992proceedings,deriglazov2011reparametrization}: reconciling how time enters the Schrödinger equation as an essentially classical, independent, and absolute parameter $t$; whereas in relativity, time has a relative connotation depending on the observer and distribution of mass-energy.  Attempts to also associate time with measurements of a \emph{time-operator} have been criticised by Dirac based on the Stone-von Neumann theorem \cite{Kronz2005} because the variable that is naturally conjugate to time is energy. And if the Hamiltonian operator $\hat H$ and a time operator $\hat T$ were to satisfy the canonical commutation relations $[\hat T, \hat H] = i$, then the eigenvalue spectrum of the Hamiltonian (and thus the energy) would need to be unbounded (from above and below). 
\\
\\
 Why would a time operator be desirable? In general relativity, as well as in more general theories with reparametrization invariance \cite{deriglazov2000local, deriglazov2009improved, deriglazov1995notes}, there is only a soft distinction between time and space. Any foliation of spacetime into non-intersecting space-like hypersurfaces defines a global time coordinate, or slicing of spacetime, which allows us to describe the evolution of a physical system in terms of a well-defined ``time'' parameter. The separation of two spacetime events can be space-like even if they are separated by a non-zero amount of coordinate time. This (literal) relativity in the definitions of time and space leads, in the canonical quantization of general relativity, to the Wheeler–DeWitt equation \cite{dewitt1967quantum}, which is based on \emph{constraints} of the form $\hat J \ket{\Psi} = 0$~, where $\hat J$ is a \emph{constraint operator} \cite{dirac1950generalized, dirac1988lectures, dewitt1967quantum} (see also \cite{Isham:1992ms, anderson_problem_time, kuchavr1992proceedings} for a discussion of the closely-related issue of the ``problem of time'' in quantum gravity). Physical states $\ket{\Psi}$ are required to satisfy the constraint (that is, are annihilated by it) and are hence in the $0$-eigenspace of the constraint operator. They represent configurations of quantum gravity degrees of freedom that describe (superpositions of) full 3+1 dimensional spacetimes. In such a setting there is no notion of a Schrödinger-like time parameter, and time should instead be emergent, such as being a measurement conditional on a specified event.
\\
\\
A possible resolution to these challenges is to forgo treating time as a fundamental external parameter and instead describe its emergence through correlations among quantum subsystems. A well-known quantum formulation of this relational perspective is the Page-Wootters (PW) framework \cite{page1983evolution, wootters1984time}, which demonstrates how conventional quantum time evolution can arise from the entanglement between two subsystems within a globally stationary quantum state. In this framework, the global quantum state is constrained to be static, and the apparent flow of time is recovered through conditional evolution relative to the clock’s degrees of freedom. The PW formalism provides an elegant resolution to the problem of time in non-interacting quantum systems, but a complete description of gravitational effects requires extending the framework to incorporate mass-energy interactions. 
\\
\\
Our work builds on the Page-Wootters formalism and is inspired by recent advancements in studies of relational time, quantum clocks, and quantum reference frames (see \cite{Smith:2019imm, Giacomini_2021, Rijavec_2023, Paczos_2024} for some representative papers, and references therein), we demonstrate that gravitational interaction and time dilation can arise naturally from quantum constraints, reinforcing the idea that gravity may emerge from fundamental quantum principles. We extend the Page-Wootters relational time formalism by introducing a mass-energy coupling to coordinate time, allowing gravitational effects to emerge naturally within a fully quantum setting. Specifically, we work with the Hamiltonian constraint equation, akin to the Wheeler-DeWitt equation, and introduce an interaction between mass-energy and the global time observable. This modification leads to two key results: (i)  the emergence of an effective potential akin to the Newtonian gravitational potential in the low-energy limit when multiple quantum systems independently and universally couple to the coordinate time degree of freedom, and (ii) a quantum analog of gravitational time dilation in the Schwarzchild metric, which matches leading-order predictions from general relativity. These results hinge on taking the interaction strength to be representative of the gravitational coupling $G$. We emphasize that the Page–Wootters construction employed here is not intended as a full quantization scheme for general relativity. The formalism is tailored to yield a Schrödinger-type evolution for subsystems, which is appropriate in the weak-field/low-curvature regime where non-relativistic quantum mechanics is a good approximation to curved spacetime dynamics. Our results should therefore be interpreted in this domain of validity. In strongly relativistic or cosmological settings, where no global or uniform time exists, a more general framework would be required.
\\
\\
The structure of this paper is as follows: In Section 2, we briefly review the Page-Wootters formalism of relational time, and use it to introduce the concept of coordinate time as a relational quantum observable and further describe how proper time of a particle can be encoded. Section 3 develops a mass-energy coupling mechanism by introducing an interaction term in the constraint leading to a modified generator of time translations which can be interpreted as time evolution under an effective Hamiltonian. In Section 4, we show how a Newtonian gravitational potential emerges from this universal coupling with the coordinate clock in the weak-field limit. In Section 5, we derive gravitational time dilation for a Schwarzchild metric-like setup, demonstrating consistency with the predictions of general relativity. Finally, we discuss implications and future directions in Section 6, outlining experimental prospects and potential extensions to quantum gravity frameworks.

\section{Relational Time, Quantum Clocks, and the Page-Wootters Formalism}
\subsection{Coordinate Time as a Relational Quantum Observable}\label{sec:coord_time_QM}
We begin by reviewing (an extension of) the Page-Wootters construction \cite{page1983evolution,wootters1984time,page1989itp} (see also related ideas by Dirac \cite{dirac1950generalized,dirac1988lectures}) which will serve as a simplified, albeit representative version of treating time as a bona fide internal (relational) quantum degree of freedom in an otherwise ``timeless'' universe. The global quantum state is static and the apparent ``flow" of time is due to the entanglement and correlations between the temporal degree of freedom with the rest of Hilbert space. We will closely
follow the work in Ref. \cite{giovannetti2015quantum} for this quick review. Here, the global Hilbert space $\hs$ is factorized into a temporal degree of freedom $\hst$, often called the \emph{clock,} and the system $\hs_{S}$ (what we typically describe in conventional quantum mechanics),
\begin{equation}
\label{eq:PW_HS}
\hs \: \simeq \: \hst \: \otimes \: \hs_{S} \: .
\end{equation}
The temporal Hilbert space $\hst$ is taken isomorphic to $\mathbb{L}_{2}(\mathbb{R})$ and we consider a pair of Hermitian operators $\toptr$ and $\pt$ on that space which satisfy Heisenberg canonical commutation relation (CCR), $\lcb \toptr , \pt \rcb = i$ 
(in units with $\hbar = 1$). These operators thus constitute a standard pair of conjugate operators on $\hst$. In particular, $\toptr$ has a continuous spectrum of eigenvalues $t$ with corresponding orthonormal eigenstates $\ket{t}$ with $\braket{t'|t} = \delta(t -t')$, such that the conjugate operator $\pt$ generates translations in $t$ as $e^{-i\pt t'}\ket{t} = \ket{t + t'}$.
Within the Page-Wootters formalism the states $\ket{t}$ are endowed with a temporal meaning by introducing the constraint operator $\J$,
\begin{equation}
\label{eq:PW_J}
\J \: = \: \pt \otimes \eye_{S} + \eye_{t} \otimes \ham_{S} \: ,
\end{equation}
where $\eye_{t}$ and $\eye_{S}$ are identity operators on $\hst$ and $\hs_{S}$, respectively, and $\ham_S$ is the conventional Hamiltonian for the system. Physical states $\kett{\Psi}$ in the global Hilbert space $\hs$ are identified to be the ones annihilated by the constraint operator $\J$, akin to the Wheeler-DeWitt equation,
\begin{equation}
\J \: \approx \: 0 \: \implies \: \J \: \kett{\Psi}\: = \: 0 \: .
\end{equation} 
We use the double-ket notation $\kett{\Psi}$ to stress that the state is defined on the global Hilbert space $\hst\otimes \hs_S$. Such a technique of quantization based on constraints can be attributed to Dirac \cite{dirac1950generalized,dirac1988lectures}. These physical states, which are eigenstates of the constraint operator $\J$ with eigenvalue zero are globally static but encode an apparent flow of time from the perspective of $\hs_S$. Conventional time-dependent states of the system are obtained via a relational approach by conditioning the global, physical state $\kett{\Psi}$ with the eigenvector $\ket{t}$ of the time operator $\toptr$,
\begin{equation}
\ket{\psi(t)} \: = \:\big\langle t \kett{\Psi} \: \in \hs_S \: .
\end{equation}
The constraint equation conditioned on $\ket{t}$ becomes
\begin{equation}
\label{eq:PW_Schrodinger}
\bra{t}\J\kett{\Psi} \: = \: \bra{t}\pt\otimes\eye_{S} \kett{\Psi} \: + \: \ham_{S}\ket{\psi(t)} \: = 0 \: ,
\end{equation}
which upon inserting a complete set of states on $\hst$ given by $\int dt \: \ket{t}\bra{t} = \eye_{t}$, and remembering that the matrix elements of the conjugate momenta are $\braket{t | \pt | t'} = -i \frac{\partial}{\partial t} \delta(t - t')$, gets us the time evolution equation for states $\ket{\psi(t)}$ of the system,
\begin{equation}
\ham_{S} \ket{\psi(t)} \: = \: i \frac{\partial}{\partial t} \ket{\psi(t)} \: .
\end{equation}
This is indeed the Schr\"odinger equation for $\ket{\psi(t)} \in \hs_S$. Thus we see that effective time evolution for states in the subfactor $\hs_S$~ of the global Hilbert space, governed by a Hamiltonian $\ham_S$~, can be recovered from a constraint operator. The Page-Wootters formalism also offers an interesting take on Pauli's criticism \citep{Leon_2017} on treating time as an operator: following the Schr\"odinger equation, one might wish to establish a conjugate relationship between the Hamiltonian and the time operator as canonically conjugate variables but this is not allowed due to the Stone-von Neumann theorem. The theorem demands a set of conjugate operators satisfying the Heisenberg CCR to have their eigenvalue spectra unbounded from below; but for physical theories, the Hamiltonian has a ground state with an energy bounded from below. In the Page-Wootters construction, the time operator $\toptr$ and the system Hamiltonian $\ham_S$ are not conjugates since they act on different Hilbert spaces. There is a bona fide pair of conjugate operators on $\hst$, the time operator $\toptr$ and its conjugate $\pt$ which satisfy the Heisenberg CCR. One can, however, argue that this is simply shifts the Pauli objection about the unbounded Hamiltonian from the system to the clock which in relational quantum mechanics is treated as another physical system with a bounded spectrum. But we note that realistic clocks will always have a finite resolution and period \citep{BUSCH1994357}, for which we can use finite-dimensional Hilbert spaces (such as discussed in refs \cite{Loveridge_2019, Singh:2018qzk}) to model such canonical conjugate variables, or use POVMs \cite{PhysRevD.104.066001, Favalli2020timeobservablesin} to construct a regularized version of the clock. Realistic clock Hamiltonians will thus still be bounded. But we will set aside this technicality in the following and continue to assume ideal clocks.
\\
\\
One can furthermore show that any physical state can be expressed as
\begin{equation}
\label{eq:global_state_v2}
    \kett{\Psi} = \int dt \ket{t} \otimes \ket{\psi(t)} \: = \int dt \ket{t} \otimes \left( e^{-i\ham_{S}t} \ket{\psi(0)} \right)\: ,
\end{equation}
where $\hat{U}(t) = e^{-i\ham_{S}t}$ is the time evolution operator acting on an initial state $\ket{\psi(0)}$ of the system.  Time evolution is hence implemented in the Page-Wootters construction via entanglement between the time Hilbert space and the system Hilbert space. In the following we will think of the degree of freedom in $\hs_t$ as representing \textit{coordinate time.} In other words, we take it to represent a very simplified version of the spacetime degrees of freedom that appear e.g.\ in the Wheeler-DeWitt equation. A key observation in this relational time formalism is that the global Wheeler-DeWitt-like constraint of Eq. (\ref{eq:PW_J}) does not couple the clock with the system via interactions, \emph{i.e.}, it treats them as non-interacting subsystems. This is acceptable when dealing with non-gravitational systems, but will have to be modified when dealing with general relativistic systems where the backreaction from mass-energy on the spacetime will be taken into account.

\subsection{Proper Time as the ``System'' Observable}

While the coordinate time represents a global foliation of spacetime and the choice of observer's coordinates, the \emph{proper time} experienced by a physics system $\hs_S$ is, in general, different than the coordinate time, and will be governed by the behavior of clocks carried by that system. To this end, we associate the system to have internal structure which serves as a clock to measure its proper time (imagine the system carrying a clock with it)\cite{Smith:2019imm}. Since we are focused on temporal effects in this work, we will only consider such internal proper clocks, and we neglect spatial motion of their center of mass. On the system Hilbert space $\hs_S$, we can associate a proper time observable $\hat{\tau}_{S}$ and its corresponding conjugate $\hat{\pi}_{S}$ which satisfies Heisenberg canonical commutation relation, $\lcb \hat{\tau}_{S} , \hat{\pi}_{S} \rcb = i$. The Hamiltonian of the system then consists of its rest mass $(m)$ energy, and a term $\hat{H}^{\rm{clock}}_{S}$ which drives evolution of the internal proper time observable (we work with units of $c = 1$),
\begin{equation}
    \ham_{S} = m \:\eye_{S} + \hat{H}^{\rm{clock}}_{S} \: .
\end{equation}
For simplicity, we take the internal proper time degree of freedom to be an ideal clock\footnote{Operationally, one can use a regularized model using finite-dimensional Hilbert spaces or POVMs to overcome Pauli's objection as we discussed in Sec. \ref{sec:coord_time_QM}} where different proper times are completely distinguishable, which leads the Hamiltonian to be the conjugate operator $\hat{\pi}_S$ that drives translation of the proper time observable,
\begin{equation}
    \hat{H}^{\rm{clock}}_{S} = \hat{\pi}_{S} \: .
\end{equation}
 With the above system in mind, the Page-Wootters constraint operator takes the following form,
\begin{equation}
    \label{eq:PW_S_clock}
    \J_{0} \: = \: \pt \otimes \eye_{S} + \eye_{t} \otimes \left(m \: \eye_{S} + \hat{\pi}_{S}\right)  \: \approx \: 0 \: ,
\end{equation}
for which the physical states take the form of Eq (\ref{eq:global_state_v2})
with the system Hamiltonian being $\ham_S = m \: \eye_{S} + \hat{\pi}_{S}$. We indicate the non-interacting nature of the coordinate time with the quantum clock system by denoting the constraint as $\J_0$.
\\
\\
To probe the relationship between coordinate time $t$ and the proper time $\tau$ of the system, we consider the probability that the system's proper time $\tau_S$ reads some value $\tau$ conditioned on the coordinate time reading $t_c = t$. This approach is inspired by the analysis of Ref. \cite{Smith:2019imm} where the special relativistic time dilation in quantum clocks is considered (see, for instance, \cite{CastroRuiz2017, Castro_Ruiz_2020} and references therein, for a discussion about quantum clocks in a gravitational setup). We can compute the conditional probability from the global physical state of Eq. (\ref{eq:global_state_v2}) by applying the projective measurements $\hat{E}_{t}(t) = \ket{t}\bra{t}$ and $\hat{E}_{S}(\tau) = \ket{\tau}_{S}\bra{\tau}$ (operators are paired with the relevant identity operators to be able to act on the full Hilbert space $\hs$) along with the Born rule,
\begin{equation}
    \label{eq:cond_prob_no_interaction}
    \begin{split}
    \mathrm{Prob} \left[ \tau_{S} = \tau | t_c = t \right]\ &= \: \: \frac{\mathrm{Prob} \left[ \tau_{S} = \tau \mathrm{\: and \:} t_c = t \right] }{\mathrm{Prob} \left[t_c = t \right] } \: , \\
&= \: \: \frac{\braa{\Psi}  \hat{E}_{t}(t)   \hat{E}_{S}(\tau) \kett{\Psi}  }{\braa{\Psi}  \hat{E}_{t}(t)   \kett{\Psi}  } \: .
    \end{split}
\end{equation}
Since the quantum state of the system $\ket{\psi(t)} = \langle t   \kett{\Psi}$ is normalized on each time slice, which is obtained via a projective measurement with $\hat{E}_{t}(t)$, we can see that $\braa{\Psi}  \hat{E}_{t}(t)   \kett{\Psi}  = \braket{\psi(t) |{\psi(t)}}  = 1 \:, \: \forall \: t$. This allows us to write the conditional probability of Eq. (\ref{eq:cond_prob_no_interaction}) as,
\begin{equation}
    \label{eq:cond_prob_no_interaction_rewrite}
    \begin{split}
    \mathrm{Prob} \left[ \tau_{S} = \tau | t_c = t \right] \: \: = \: \: \bra{\psi(t)}   \hat{E}_{S}(\tau) \ket{\psi(t)}   \: .
    \end{split}
\end{equation}
Using unitary time evolution, $\ket{\psi(t)} = \hat{U}(t) \ket{ \psi(0)}$, the conditional probability can be written as matrix elements of the time evolution operator acting on the initial system state,
\begin{equation}
\label{eq:cond_prob_step3}
\begin{split}
    \mathrm{Prob} \left[ \tau_{S} = \tau | t_c = t \right] \: \:&=\: \: \braket{\psi(0)  | \hat{U}^{\dag}(t) \hat{E}_{S}(\tau) \hat{U}(t) | \psi(0)} \: , \\
    &= \braket{\psi(0)  | \hat{U}^{\dag}(t) \ket{\tau}_{S}\bra{\tau} \hat{U}(t) | \psi(0)} \: , \\
&=  \:\: \big|  \braket{\tau_{S} = \tau    | e^{-i\ham_{S} t} |\psi(0) }\big|^{2} \: .
    \end{split}
\end{equation}
We take the initial ($t = 0$) state $\ket{\psi(0)}$ of the internal proper clock to be a fiducial Gaussian wave packet centered around $\tau = 0$ with a width of $\sigma$ in the eigenbasis of the proper time observable $\hat{\tau}_S$, 
\begin{equation}
\label{eq:init_state_system_clock}
    \ket{\psi(0)} = \frac{1}{\left(2 \pi \sigma^{2}\right)^{1/4}} \int d\tilde{\tau} \: e^{-\frac{\tilde{\tau}^{2}}{4\sigma^{2}}} \: \ket{\tilde{\tau}}  \: .
\end{equation}

The action of $ e^{-i\ham_{S} t}$ on $\ket{\psi(0)}$ can be computed by using the fact that $\hat{\pi}_{S}$ generates translations in eigenstates of $\hat{\tau}_S$, as $e^{-i\hat{\pi}_{S}t}\ket{\tilde{\tau}} = \ket{\tilde{\tau} + t}$,
\begin{equation}
    e^{-i\ham_{S} t}\ket{\psi(0)} = \frac{e^{-imt}}{\left(2 \pi \sigma^{2}\right)^{1/4}} \int d\tilde{\tau} \: e^{-\frac{\tilde{\tau}^{2}}{4\sigma^{2}}} \: \ket{\tilde{\tau} + t}  \: ,
\end{equation}
which upon taking an inner product with $\ket{\tau_{S} = \tau}$ and remembering $\braket{\tau | \tilde{\tau} + t} = \delta(\tilde{\tau} - (\tau - t))$, our desired conditional probability simplifies to,
\begin{equation}
\label{eq:cond_prob_no_int}
    \mathrm{Prob} \left[ \tau_{S} = \tau | t_c = t \right] \: \: = \: \: \frac{1}{\sqrt{2\pi \sigma^{2}}} e^{- \frac{(\tau - t)^{2}}{2\sigma^{2}}  } \: .
\end{equation}
We can now use this conditional probability distribution to compute the average proper time as read by the system $S$ conditioned on the coordinate time reading $t$. Given the probability distribution of Eq. (\ref{eq:cond_prob_no_int}) is a Gaussian, the average coincides with the mode of the distribution, in this case, it is at the coordinate time reading $t_c = t$,
\begin{equation}
\label{eq:tau_is_t}
    \langle \tau_{S} \rangle = t \: .
\end{equation}
No surprises here. On average, the proper time of the system, quantum mechanical or not, is the same as the coordinate time, as one would expect from a non-gravitational, non-relativistic setup.

\section{Mass-Energy Coupling with Coordinate Time} \label{sec:massenergy_coupling}
While the Page-Wootters framework provides a compelling relational foundation for time, it assumes that the clock and the system evolve independently, without any interaction. However, in a gravitational setting, mass-energy will backreact on spacetime, and any fundamental treatment of time must account for this backreaction. To incorporate gravitational effects, we now introduce a coupling between system (carrying mass-energy) and the global time observable. This modification leads to a deformation of the system’s time evolution, revealing key features such as gravitational time dilation and the emergence of an effective gravitational potential.
\\
\\
To express this backreaction in a simplified representative model, we introduce an \textit{interaction term} between the coordinate clock and the system (see also \cite{smith2019quantizing} for a similar idea) at the level of the Wheeler-DeWitt constraint,
\begin{equation}
\label{eq:J_interacting}
    \J \: = \: \pt \otimes \eye_{S} + \eye_{t} \otimes \ham_{S} + \frac{1}{\Lambda}\left(\pt \otimes \ham_{S}\right) \: \approx \: 0 \: ,
\end{equation}
where we have added a term $\left(\pt \otimes \ham_{S}\right)/\Lambda$ to the standard Page-Wootters constraint of Eq. (\ref{eq:PW_J}) to model gravitational back reaction of the system on the coordinate degree of freedom $\hs_t$. Two points are worth noting here. First, the form of the interaction is gravitationally motivated, that is, the rationale for our form of the interaction term comes from linearized quantum gravity which couples first order energy terms to linear metric perturbations. In general, one could motivate such terms by exploring order-by-order terms in the Hamiltonian constraint of classical general relativity. While the sign of the coupling constant is arbitrary at this state, we will later see in Section \ref{sec:newton_grav} that a positive coupling $\Lambda > 0$ will allow us to match results with those of classical general relativity to leading order. Second, the interaction commutes with the self-Hamiltonian, as has been argued by Ref \cite{Marletto_2022} which constraints the form of additive terms to the Hamiltonian. The exact form of this interaction term will be sensitive to finer details of the energy-momentum tensor and its dynamics, but we argue that our model captures the essential features needed for our discussion. We will motivate later that the interaction strength $\Lambda^{-1}$ of that term should be $\mathcal{O}(G/R)$, with $G$ being Newton's constant and $R$ being a typical spatial scale of the system. 
\\
\\
Physical states are those which are annihilated by the interacting constraint of Eq. (\ref{eq:J_interacting}).  The effect of this \textit{interaction term} is in fact equivalent to modifying the generator of translations of coordinate time (the ``momentum'' $\pt$ of the time degree of freedom). To see this, we notice that the the null space of the following operator is the same as that of $\J$ in Eq. (\ref{eq:J_interacting}),
\begin{equation}
\label{eq:PW_interacting_constraint2}
   \hat{\tilde{\mathbb{J}}}  \: = \: \pt \left( \eye_{t} + \frac{\pt}{\Lambda}\right)^{-1} \otimes \: \eye_{S} \: + \:  \eye_{t} \otimes \ham_{S} \: \approx \: 0 \: .
\end{equation}
Strictly speaking, for an ideal clock with an unbounded spectrum the operator $\left( \eye_{t} + {\pt}/{\Lambda}\right)$ is not globally invertible (a similar point is discussed in ref. \citep{Rijavec_2023}). In this work, we implicitly restrict attention to the perturbative regime with weak coupling (as would be made clear in Sections \ref{sec:newton_grav} and \ref{sec: grav_time}) where a series expansion of $\left( \eye_{t} + {\pt}/{\Lambda}\right)^{-1}$ is valid, or use to finite-dimensional clock models where the spectrum is bounded and such that the inverse is well-defined. Alternatively, one can exclude the eigensector corresponding to the clock momenta which lead to divergences, though this may lack physical motivation. A complete analysis of the divergence that would occur in the full non-perturbative regime is left for future work. The set of physical states satisfying the constraint of Eq. (\ref{eq:J_interacting}) is also the set of states satisfying the constraint of Eq. (\ref{eq:PW_interacting_constraint2}), that is $\J \kett{\Psi} = 0 \: \implies \:  \hat{\tilde{\mathbb{J}}} \kett{\Psi} = 0$. Thus, we see that the interacting constraint (representative of coupling energy with the background metric) can be cast into an equivalent non-interacting constraint with an effectively modified rate of flow of coordinate time.
Since the interaction term commutes with the standard Page-Wootters constraint of Eq. (\ref{eq:PW_J}), 
\begin{equation}
    \lcb \left(\pt \otimes \eye_{S} + \eye_{t} \otimes \ham_{S} \right) \: \: ,\: \:   \frac{1}{\Lambda}\left(\pt \otimes \ham_{S}\right) \rcb = 0 \: ,
\end{equation}
we can construct special separable, tensor product\footnote{For simplicity, and without loss of generality, we assume that the eigenspectrum of the system Hamiltonian is non-degenerate.} eigenstates of the form
\begin{equation}
    \kett{\Psi}_{(\alpha,n)} =  \:\:\ket{p_{t} = \alpha }  \ket{E_n} \: ,
\end{equation}
where $\ket{p_{t} = \alpha}$ is an eigenstate of the unmodified conjugate operator $\pt$ of coordinate time with eigenvalue $\alpha \in \mathbb{R}$, and $\ket{E_n}$ is an eigenstate of the system Hamiltonian $\ham_{S}$ with eigenvalue $E_n$ for some index\footnote{We are labeling the energy eigenvalues of $\ham_{S}$ by a discrete index $n$ for convenience, but we can work with a continuous spectrum just as well.} $n$. In general, eigenstates of $\hat{\mathbb{J}}$ will be a superposition of the subset of such separable state which share the same eigenvalue, hence leading to entangled eigenstates. For these eigenstates to be physical, they must be annihilated by the constraint of Eq. (\ref{eq:J_interacting}) (or equivalently, Eq. (\ref{eq:PW_interacting_constraint2})), which then constrains the allowed values of coordinate clock momentum in terms of energy of the system. For instance, for each of the allowed separable states (whose superposition will make a generic eigenstate), we have, following Eq. (\ref{eq:J_interacting}),
\begin{equation}
\label{eq:freq_time_modified}
    \alpha + E_{n} + \frac{\alpha E_{n}}{\Lambda} = 0 \: \: \:  \: \: \: \implies  \: \: \: \: \: \: \alpha_n = -\left(\frac{E_{n}}{1 + E_{n}/\Lambda}\right) \: .
\end{equation}
For the special case of the non-interacting constraint of Eq. (\ref{eq:PW_S_clock}), we would have simply obtained $\alpha_{n} = -E_{n}$ which is a spectral parallel for the one-to-one connection of proper time and coordinate time of Eq. (\ref{eq:tau_is_t}) in the case of no coupling.
\\
\\
Generic physical states (the $0-$ eigenspace of $\hat{\mathbb{J}}$) annihilated by the modified constraint of Eq. (\ref{eq:PW_interacting_constraint2}), or equivalently by Eq. (\ref{eq:J_interacting}), then have the entangled form with $\alpha_{n}$ as derived above in Eq. (\ref{eq:freq_time_modified}),
\begin{equation}
    \kett{\Psi} = \sum_{n} c_{n} \ket{p_{t} = \alpha_{n} }  \ket{E_n} \: ,
\end{equation}
which written in the conjugate time basis will yield,
\begin{equation}
    \kett{\Psi} = \int dt \: \ket{t} \otimes \left(\sum_{n} c_{n} e^{-i\left(\frac{E_n}{1 + E_{n}/\Lambda}\right)t}\right) \ket{E_{n}}   \: .
\end{equation}
 At the level of the static global quantum state where time evolution is interpreted as relational between the clock and the system, this allows an equivalent representation which expresses (coordinate) time evolution of the system under a \emph{modified} Hamiltonian,
\begin{equation}
\label{eq:modified_U_evolution}
    \kett{\Psi} = \int dt \: \ket{t} \otimes \left(e^{-i\hat{H}_{\mathrm{eff}}t} \ket{\psi(0)}\right) \: ,
\end{equation}
where $\hat{H}_{\mathrm{eff}}$ is the effective Hamiltonian of the system's evolution induced by interaction of the system with the coordinate clock,
\begin{equation}\label{eq:eff_Hamiltonian}
    \hat{H}_{\mathrm{eff}} = \ham_{S} \left(\eye_{S} + \frac{\ham_{S}}{\Lambda}\right)^{-1} \: .
\end{equation}
Recall that the energy dependent phase, ``$\exp{(-iEt)}$,'' in quantum mechanics is responsible for time evolution. The coupling between system energy and background metric's coordinate time led to a modified generator of (coordinate) time translations, which in turn can be interpreted as time flowing at different rates for states with different energy content, as is a key feature of gravitation. 
\section{Emergent Newtonian Gravitational Potential}\label{sec:newton_grav}


In linearized classical general relativity, two masses backreacting on spacetime lead to an effective Newtonian gravitational potential between them, to leading order. We now explore how a similar Newtonian-like potential can emerge in the quantum relational setup of Sec. \ref{sec:massenergy_coupling} by coupling mass-energy to the coordinate time. In particular, the interacting constraint of Eq. (\ref{eq:J_interacting}) alters the effective Hamiltonian governing system evolution as seen in Eq. (\ref{eq:eff_Hamiltonian}), introducing interaction terms between subsystems of $\hs_S$. Remarkably, in the low-energy limit, this interaction term between the system and the coordinate clock manifests itself akin to an emergent Newtonian gravitational potential between subsystem of $\hs_S$. This suggests that gravitational attraction can arise naturally within a purely quantum framework, without being explicitly imposed. We now explore this emergent gravitational interaction in detail.
\\
\\
 To do this, we take our system $\hs_S$ to consist of two \textit{non-interacting} particles $A$ and $B$ with rest masses $m_{A}$ and $m_{B}$, respectively, held at a constant distance from each other. The system Hilbert space $\hs$ is now decomposed as,
\begin{equation}
    \hs_S \simeq \hs_{A} \otimes \hs_{B} \: ,
\end{equation}
where $\hs_{A}$ and $\hs_{B}$ represent the Hilbert spaces corresponding to the internal degrees of freedom of the two particles, respectively. In our interacting constraint model of Sec. \ref{sec:massenergy_coupling}, the total hamiltonian of the system is coupled with the coordinate clock, with $\Lambda$ controling the coupling strength of the full system with the clock. In the particular case where the system consists of two non-interacting particles (which we study here, it leads to each subsystem coupling with the common coordinate-time degree of freedom, in analogy with gravitational universal coupling to energy through the spacetime metric. The resulting cross term then plays the role of an emergent Newtonian potential between the systems, arising through the shared background clock rather than being imposed directly. In case other forces between the susbsystems have to be included, such as electromagnetic interaction, those can be added as an interaction between $\hs_A$ and $\hs_B$. Here, we only focus on studying gravitational-like interaction mediated through coupling independently with the coordinate clock.

 To take into account their relative motion, we would need a proper quantum treatment for space as well. In this work, we are focusing exclusively on time, and we leave a similar quantum framework for spatial degrees of freedom for future work. In particular, we associate each particle to have internal structure which serves as a ``clock'' it carries with it to measure its proper time. On each of these Hilbert spaces, we can associate a proper time observable and its corresponding conjugate momentum satisfying Heisenberg canonical commutation, for $A$, $\lcb \hat{\tau}_{A} , \hat{\pi}_{A} \rcb = i$, and for $B$, we have $\lcb \hat{\tau}_{B} , \hat{\pi}_{B} \rcb = i$. The Hamiltonian for this non-interacting system of two particles is then simply,
\begin{equation}
    \ham_{AB} = \hat{h}_{A} \otimes \eye_{B} + \eye_{A} \otimes \hat{h}_{B} \: \equiv \: \ham_{A} + \ham_{B} \: ,
\end{equation}
where the self-Hamiltonians of the two particles consist of their rest mass energy and a term $\hat{H}^{\rm{clock}}$ which drives evolution of their internal proper time observable,
\begin{equation}
    \hat{h}_{A} = m_{A} \eye_{A} + \hat{H}^{\rm{clock}}_{A} \:, \: \: \: \:\: \: \mathrm{and}  \: \:\: \:\: \: \hat{h}_{B} = m_{B} \eye_{B} + \hat{H}^{\rm{clock}}_{B} \: .
\end{equation}
For ideal clocks, as discussed before, we take the Hamiltonian of the proper time evolution of each particle to be given by the corresponding conjugate $\hat{\pi}$
\begin{equation}
    \hat{H}^{\rm{clock}}_{j} = \hat{\pi}_{j} \: , \: \:\:\:\:\:\: j \in \{A,B\} \: .
\end{equation}
The \textit{interacting} Page-Wootters constraint of Eq. (\ref{eq:J_interacting}) with this two-body system now takes the form,
\begin{equation}
    \label{eq:PW_2_particles}
    \J \: = \: \pt \otimes \eye_{AB} + \eye_{t} \otimes \left(\ham_{A} + \ham_{B} \right) + \frac{1}{\Lambda}\left(\pt \otimes (\ham_{A} + \ham_{B})\right) \: ,
\end{equation}
which as we discussed in Eq. (\ref{eq:modified_U_evolution}) leads to physical states governed by a modified system Hamiltonian, in this case,
\begin{equation}
     \hat{H}_{\mathrm{eff}} = \ham_{AB} \left(\eye_{AB} + \frac{\ham_{AB}}{\Lambda}\right)^{-1} \: .
\end{equation}
If the interaction strength $\Lambda^{-1}$ is weak, as is relevant for the interesting regime of weak-field gravity, the modification to the Hamiltonian can be retained at leading order as a good approximation,
\begin{equation}
    \hat{H}_{\mathrm{eff}} \approx \ham_{AB} \left(\eye_{AB} - \frac{\ham_{AB}}{\Lambda}\right) \: .
\end{equation}
When expressed in terms of the individual particle Hamiltonians, this takes the following interesting form:
\begin{equation}
\label{eq:H_AB_gravity}
    \hat{H}_{\mathrm{eff}} = \left(\ham_{A} - \frac{\ham_{A}^{2}}{\Lambda}\right)  + \left(\ham_{B} - \frac{\ham_{B}^{2}}{\Lambda}\right) - \frac{2}{\Lambda}\left(\ham_{A}\otimes \ham_{B}\right) \: .
\end{equation}
The modified Hamiltonian not only alters the self-Hamiltonians of the two particles, but more importantly, introduces a \emph{purely quantum-mechanical interaction} term which between the two particles. The form of the interaction matches a Newtonian gravitational potential energy under the following identification,
\begin{equation}
    \frac{1}{\Lambda} = \frac{G}{2r} \: ,
\end{equation}
$r$ being the constant (classical) separation between the two particles. We make it clear that the form is indicative of the Newtonian gravitational interaction but a full identification cannot yet be made since we have not fully dealt with spatial degrees of freedom in this work. We emphasize that in our present framework the spatial separation $r$ is treated as an external parameter introduced when matching the effective interaction term with a Newtonian potential. This arises because, at this stage, the coupling is only between each system and the common coordinate-time degree of freedom. In contrast, in approaches such as ref \citep{CastroRuiz2017}, a spatial coordinate system is assumed to exist and clock behavior is chosen to be consistent with mass-energy equivalence in general relativity, thereby directly giving the distance dependence in the gravitational potential. Our result should therefore be viewed as establishing that a universal clock-mediated coupling naturally produces an $H_{A} H_{B}$-type interaction, whose coefficient can then be identified with the Newtonian case at leading order. Including spatial degrees of freedom explicitly in such a Page-Wootters-like setup (such as \citep{Singh:2020kdu}), one may expect $\Lambda$ to acquire the interpretation of an effective coupling parameter encoding the separation between the systems. A more complete analysis of this distance dependence is left for future work.

 We do, however, show that under this choice of $\Lambda$, we also recover gravitational time dilation in a Schwarzchild metric to leading order, as we show in Section (\ref{sec: grav_time}), lending credence to this choice. We note that the choice of a positive $\Lambda$ in the constraint is essential in obtaining the effective interaction with the correct (negative) sign, corresponding to an attractive Newtonian potential and in agreement with classical GR at leading order. In contemporary works such as ref. \citep{CastroRuiz2017}, the same sign emerges directly from mass–energy equivalence and the negative Newtonian potential. In our Page–Wootters construction, the positive coupling in the constraint leads, through the constraint enforcing opposite shifts of the clock and the system, to the required negative sign in the effective Hamiltonian coupling. It is worth emphasizing that this Newtonian gravitational-like interaction has emerged as a consequence of interaction added at the level of the Wheeler de-Witt constraint between the coordinate clock and the system of particles. No direct, a-priori interaction was introduced between the particles, but it has emerged as a consequence of matter universally interacting with the coordinate clock. Note, in particular, that the charges of the emergent force are the self-Hamiltonians of the particles, as is appropriate for gravity under the mass-energy equivalence.

\section{Gravitational Time Dilation}\label{sec: grav_time}

Following the emergence of an effective Newtonian gravitational potential from the interacting constraint, we now turn to its direct consequence on time evolution. In general relativity, gravitational fields alter the passage of time, leading to gravitational time dilation, such as that described by the Schwarzschild metric. In our framework, a similar effect arises from the quantum mechanical interaction between mass-energy and coordinate clock. This modifies the system’s proper time evolution, yielding a quantum gravitational time dilation effect. We now derive this result and compare it with the classical predictions of general relativity.
\\
\\
To do so, we investigate how the proper time as measured by a particle correlates with the coordinate time when mass-energy back reacts on the metric (as modeled by the interacting constraint) representing a gravitational field? To do so, we calculate the conditional probability of the test particle reading a certain proper time $\tau$ given the coordinate time reads $t$. Our physical states now take the following form,
\begin{equation}
\label{eq:AB_state_noninteracting}
    \kett{\Psi} = \int dt \ket{t} \otimes  \left( e^{-i\ham_{\rm{eff}}t}\ket{\psi_{AB}(0)} \right)\: ,
\end{equation}
where $\ham_{\rm{eff}} $ is the leading order modified Hamiltonian for the two-particle system given in Eq. (\ref{eq:H_AB_gravity}) as induced by the interacting constraint. 
To make contact with the standard Schwarzchild metric in general relativity as our working example, we will work in context of two objects, one serving as a massive body, and the other as a test particle reacting to the metric as it responds to the energy of the massive body. Consider particle $A$ to be a highly massive object with mass $m_{A} = M$, but with no internal structure, \textit{i.e.} $\ham_{A}^{\rm{clock}} = 0$. One can neglect the internal clock structure of $A$ on the grounds that its rest mass energy dominates over any other energy scales involved in its spectrum in accordance with its role of the stationary, central mass. Particle $B$ serves as a test mass which probes proper time dilation effects due to gravity. It therefore has a negligibly small rest mass $m_{B} = m << M$ so that is does not significantly back-react on the coordinate clock, but only responds to the back-reaction created by object $A$. With this setup, the bare self-Hamiltonians of the two particles become,
\begin{equation}
    \hat{h}_{A} = M  \:, \: \: \: \:\: \: \mathrm{and}  \: \:\: \:\: \:  \hat{h}_{B} =  m\eye_{B} +  \hat{\pi}_{B} \: ,
\end{equation}
and the interaction-induced modified Hamiltonian has the following relevant terms which act on particle $B$, 
\begin{equation}
\label{eq:Hamiltonian_B_modified}
\hat{H}_{\mathrm{eff}}^{(B)} =       \left(1 - \frac{2(M+m)}{\Lambda}\right)\hat{\pi}_{B} - \frac{\hat{\pi}^{2}_{B}}{\Lambda} \: + \: \left(\mathrm{mass \: dependent \: constant}\right) \: ,
\end{equation}
where the mass-dependent constant will only add an overall, irrelevant phase to the quantum state.
To probe the relationship between proper time of the test mass $B$ and the background (metric) coordinate time, we consider the probability that the proper time of particle $B$ reads some value $\tau$ conditioned on the coordinate time reading $t_c = t$. 
Similar to the calculation done in  Eqs. (\ref{eq:cond_prob_step3}) - (\ref{eq:cond_prob_no_int}), the conditional probability takes the form,
\begin{equation}
\begin{split}
    \mathrm{Prob} \left[ \tau_{B}  = \tau | t_c = t \right] &= \: \: \braket{\psi_B(t) |  \hat{E}_{B}(\tau) | \psi_B(t) } \\
    &= \:\: \big|  \braket{\tau_{B} = \tau    | e^{-i\ham^{(B)}_{\rm{eff}} t} |\psi_{B}(0) }\big|^{2} \: ,
    \end{split}
\end{equation}
where $\ham^{(B)}_{\rm{eff}}$ is the modified Hamiltonian on the test particle given in Eq. (\ref{eq:Hamiltonian_B_modified}). We take the initial quantum state (i.e.\ at $t=0$ coordinate time) of the test particle $B$ (or rather, of its internal proper clock) to be a fiducial Gaussian wave packet centered around $\tau_B = 0$ with a width of $\sigma$ in the eigenbasis of the proper time observable, $\hat{\tau}_B$. Since object $A$ is taken to be a highly massive object lacking internal quantum structure, we take its quantum state to be in an unknown, albeit fixed state which does not affect our results, and thus can be neglected. Therefore, the initial quantum state of the system $B$ is,
\begin{equation}
    \ket{\psi(0)}_{B} = \frac{1}{\left(2 \pi \sigma^{2}\right)^{1/4}} \int d\tilde{\tau} \: e^{-\frac{\tilde{\tau}^{2}}{4\sigma^{2}}} \: \ket{\tau_{B} = \tilde{\tau} }
\end{equation}

The conditional probability can be obtained by noting that the $\hat{\pi}_B$ term in the effective Hamiltonian shifts the $\hat{\tau}_B$ eigenstates, and the $\hat{\pi}^{2}_B$ term makes the wave packet spread out in time,
\begin{equation}
    \mathrm{Prob} \left[ \tau_{B} = \tau | t_c = t \right] \: \: = \: \: \frac{1}{\sqrt{2\pi \sigma(t)^{2}}} \exp{\left(- \frac{\left[  \tau - t\left(1 - \frac{2M^2}{\Lambda}\right)      \right]^{2}}{2\sigma(t)^{2}}  \right) } \: .
\end{equation}
where $\sigma(t) = \sigma\sqrt{1 + \left(\frac{t}{\sigma^{2}\Lambda}\right)^{2}}$ is now a time-dependent variance of the Gaussian probability distribution. Here, we have explicitly taken the limit $M>>m$ to account for the fact that particle $B$ is a test mass with a negligibly small mass compared to the central mass $M$. We now use this conditional probability distribution to compute the average proper time as read by the test mass conditioned on the coordinate time reading $t$, which yields,
\begin{equation}
    \langle \tau_{B} \rangle = t\left(1 - \frac{2M}{\Lambda}\right)  \: .
\end{equation}
This, to leading order, matches the gravitational time dilation as recorded by a stationary test particle in a Schwarzchild metric with a central mass $M$, under our identification $\Lambda^{-1} = G/2r$,
\begin{equation}
    d\tau = dt\sqrt{1 - \frac{2GM}{r}} \approx dt\left(1 - \frac{GM}{r}\right) \: .
\end{equation}
We note that in Section \ref{sec:newton_grav}, the same choice of $\Lambda^{-1} = G/2r$ recovered the Newtonian gravitational potential of Eq. (\ref{eq:H_AB_gravity}) between two non-interacting objects independently coupling to the coordinate clock, lending further credence to this choice. Unlike in classical relativity, where time dilation follows from the Schwarzschild metric, here it emerges as a consequence of the modified commutation relation for coordinate time. This suggests that quantum superpositions of energy states could lead to corrections beyond classical predictions, potentially observable in high-precision atomic clock experiments. With a simple model of the background coordinate time as a quantum degree of freedom coupling with mass-energy, we showed that gravitational time dilation is recovered as a consequence of a modification to the generator of coordinate time translations, as seen in Eq. (\ref{eq:PW_interacting_constraint2}),
\begin{equation}
    \hat{p}_{t,\mathrm{eff}} = \pt \left( \eye_{t} + \frac{\pt}{\Lambda}\right)^{-1}  = \: \pt \left( \eye_{t} - \frac{\pt}{\Lambda}  + \mathcal{O}\left(\frac{\pt^{2}}{\Lambda^{2}}\right)\right)\: .
\end{equation}
This not only modifies the rate of flow of coordinate time, but also induces a deformed canonical commutation relation which is a general expectation from the theory of quantum gravity, with implications such as minimal resolution, etc. (for example, \cite{Maggiore:1993rv, Ali:2009zq, Ali:2011fa}),
\begin{equation}
    \lcb \hat{t},  \hat{p}_{t,\mathrm{eff}}\rcb \: = \: i\left(\eye_t - \frac{2}{\Lambda}\pt + \mathcal{O}\left(\frac{\pt^{2}}{\Lambda^{2}}\right)\right) \: .
\end{equation}
In addition to the Newtonian gravitational potential naturally emerging in Eq. (\ref{eq:H_AB_gravity}) as a feature of this construction, we note that the modification to the Hamiltonian, including the self-term, is reminiscent of renormalization. Following Eq. (\ref{eq:freq_time_modified}), we see that even highly energetic states with $E_{n} \rightarrow \infty$ induce time evolution at a rate governed not by their bare energy, but are rather controlled by the interaction scale as $\alpha_n \rightarrow \Lambda$ in the high-energy limit, potentially softening out divergences at high energies. We note this interesting feature and aim to explore it in future work.

\section{Conclusion}

In summary, we have demonstrated that gravitational time dilation and an emergent Newtonian potential arise naturally in a quantum framework where time is treated relationally. Some of these effects have direct analogues in linearized, classical general relativity. In the weak field limit, general relativity predicts a Newtonian potential $\Phi = -GM/r$ created by a mass $M$, and the redshift of proper time $d\tau/dt \approx 1 - GM/r$. By the mass-energy equivalence, this leads to the same $H_{A}H_{B}$ Newtonian interaction and time dilation \citep{CastroRuiz2017}. We stress that this correspondence holds in the low-curvature/weak-coupling, beyond which a fully relativistic treatment would be required. We have shown that coupling otherwise non-interacting quantum systems to a global time degree of freedom leads to a number of features expected in (quantum) gravity: gravitational time dilation, agreeing with the relationship between proper and coordinate time in the Schwarzschild metric at leading order in $G$; emergence of a Newtonian potential, sourced in particular by mass-energy as is appropriate for gravity; renormalization of divergent energies. This also opens the possibility of particles in superposition back-reacting on the background coordinates leading to an additional quantum correction to gravitational time dilation, which could serve as an experimental probe of quantum gravity in the low-energy regime, such as using atom interferometers (for example, as discussed in refs. \cite{Smith:2019imm, Tino_2021}). Our findings suggest that the evolution of quantum systems, when constrained by a global time observable, inherently encodes gravitational interactions, reinforcing the idea that gravity may be a consequence of quantum constraints rather than a fundamental interaction. The implications of this work extend beyond the specific results derived here. Our approach provides a concrete realization of how gravitational effects can emerge from quantum mechanics alone, aligning with broader efforts to understand gravity as an emergent phenomenon from more fundamental quantum degrees of freedom. 
\\
\\
Several open questions remain. We focused on a specific form of the interaction in the constraint equation motivated by physically plausible arguments, but other formulations of mass-energy coupling in constrained quantization merits further exploration. A natural next step would also be to investigate whether this formulation extends to higher-order gravitational corrections and what kind of quantum contributions to it can be motivated from similar constraint-based methods. What is also missing from our considerations is a fully quantum treatment of time and space, as opposed to just time. This will be a target of future work, and we conjecture that such a combined analysis of space and time will solidify our assumption that the coupling strength $\Lambda^{-1}$ should scale as $G$ divided by a characteristic spatial scale of the system. Finally, experimental implications of quantum corrections to time dilation could be explored in high-precision atomic clock and interferometry experiments, offering a potential avenue for quantum gravity phenomenological tests in a laboratory setting.

\vspace{0.3cm}
\begin{center} 
{\bf Acknowledgments}
\end{center}

AS acknowledges funding support by the MJ Murdock Trust, and by Whitman College. OF was supported by the Excellence Cluster ORIGINS which is funded by the Deutsche Forschungsgemeinschaft (DFG, German Research Foundation) under Germany's Excellence Strategy - EXC-2094-390783311, as well as by a Fraunhofer-Schwarzschild Fellowship of the LMU Observatory.

\bibliographystyle{sn-mathphys-num}
\bibliography{revised_Found_Phys_AS_OF}


\begin{thebibliography}{37}
\ifx \bisbn   \undefined \def \bisbn  #1{ISBN #1}\fi
\ifx \binits  \undefined \def \binits#1{#1}\fi
\ifx \bauthor  \undefined \def \bauthor#1{#1}\fi
\ifx \batitle  \undefined \def \batitle#1{#1}\fi
\ifx \bjtitle  \undefined \def \bjtitle#1{#1}\fi
\ifx \bvolume  \undefined \def \bvolume#1{\textbf{#1}}\fi
\ifx \byear  \undefined \def \byear#1{#1}\fi
\ifx \bissue  \undefined \def \bissue#1{#1}\fi
\ifx \bfpage  \undefined \def \bfpage#1{#1}\fi
\ifx \blpage  \undefined \def \blpage #1{#1}\fi
\ifx \burl  \undefined \def \burl#1{\textsf{#1}}\fi
\ifx \doiurl  \undefined \def \doiurl#1{\url{https://doi.org/#1}}\fi
\ifx \betal  \undefined \def \betal{\textit{et al.}}\fi
\ifx \binstitute  \undefined \def \binstitute#1{#1}\fi
\ifx \binstitutionaled  \undefined \def \binstitutionaled#1{#1}\fi
\ifx \bctitle  \undefined \def \bctitle#1{#1}\fi
\ifx \beditor  \undefined \def \beditor#1{#1}\fi
\ifx \bpublisher  \undefined \def \bpublisher#1{#1}\fi
\ifx \bbtitle  \undefined \def \bbtitle#1{#1}\fi
\ifx \bedition  \undefined \def \bedition#1{#1}\fi
\ifx \bseriesno  \undefined \def \bseriesno#1{#1}\fi
\ifx \blocation  \undefined \def \blocation#1{#1}\fi
\ifx \bsertitle  \undefined \def \bsertitle#1{#1}\fi
\ifx \bsnm \undefined \def \bsnm#1{#1}\fi
\ifx \bsuffix \undefined \def \bsuffix#1{#1}\fi
\ifx \bparticle \undefined \def \bparticle#1{#1}\fi
\ifx \barticle \undefined \def \barticle#1{#1}\fi
\bibcommenthead
\ifx \bconfdate \undefined \def \bconfdate #1{#1}\fi
\ifx \botherref \undefined \def \botherref #1{#1}\fi
\ifx \url \undefined \def \url#1{\textsf{#1}}\fi
\ifx \bchapter \undefined \def \bchapter#1{#1}\fi
\ifx \bbook \undefined \def \bbook#1{#1}\fi
\ifx \bcomment \undefined \def \bcomment#1{#1}\fi
\ifx \oauthor \undefined \def \oauthor#1{#1}\fi
\ifx \citeauthoryear \undefined \def \citeauthoryear#1{#1}\fi
\ifx \endbibitem  \undefined \def \endbibitem {}\fi
\ifx \bconflocation  \undefined \def \bconflocation#1{#1}\fi
\ifx \arxivurl  \undefined \def \arxivurl#1{\textsf{#1}}\fi
\csname PreBibitemsHook\endcsname

\bibitem[\protect\citeauthoryear{Carroll and Singh}{2018}]{Carroll:2018rhc}
\begin{botherref}
\oauthor{\bsnm{Carroll}, \binits{S.M.}},
\oauthor{\bsnm{Singh}, \binits{A.}}:
{Mad-Dog Everettianism: Quantum Mechanics at Its Most Minimal}
(2018)
{\href{https://arxiv.org/abs/1801.08132}{{arXiv:1801.08132}}}
{[quant-ph]}
\end{botherref}
\endbibitem

\bibitem[\protect\citeauthoryear{Giddings}{2018}]{giddings2018quantum}
\begin{botherref}
\oauthor{\bsnm{Giddings}, \binits{S.B.}}:
Quantum gravity: a quantum-first approach.
arXiv preprint arXiv:1805.06900
(2018)
\end{botherref}
\endbibitem

\bibitem[\protect\citeauthoryear{Giddings}{2019}]{giddings2019quantum}
\begin{barticle}
\bauthor{\bsnm{Giddings}, \binits{S.B.}}:
\batitle{Quantum-first gravity}.
\bjtitle{Foundations of Physics}
\bvolume{49}(\bissue{3}),
\bfpage{177}--\blpage{190}
(\byear{2019})
\end{barticle}
\endbibitem

\bibitem[\protect\citeauthoryear{Isham}{1992}]{Isham:1992ms}
\begin{botherref}
\oauthor{\bsnm{Isham}, \binits{C.J.}}:
{Canonical quantum gravity and the problem of time},
0157--288
(1992)
{\href{https://arxiv.org/abs/gr-qc/9210011}{{arXiv:gr-qc/9210011}}}
{[gr-qc]}.
[NATO Sci. Ser. C409,157(1993)]
\end{botherref}
\endbibitem

\bibitem[\protect\citeauthoryear{Anderson}{2017}]{anderson_problem_time}
\begin{bbook}
\bauthor{\bsnm{Anderson}, \binits{E.}}:
\bbtitle{The Problem of Time}.
\bsertitle{Fundamental Theories of Physics, Vol 190}.
\bpublisher{Springer}, \blocation{???}
(\byear{2017}).
\burl{https://www.springer.com/us/book/9783319588469}
\end{bbook}
\endbibitem

\bibitem[\protect\citeauthoryear{Kucha{\v{r}}}{1992}]{kuchavr1992proceedings}
\begin{botherref}
\oauthor{\bsnm{Kucha{\v{r}}}, \binits{K.}}:
Proceedings of the 4th Canadian Conference on General Relativity and
  Relativistic Astrophysics.
World Scientific Singapore
(1992)
\end{botherref}
\endbibitem

\bibitem[\protect\citeauthoryear{Deriglazov and
  Rizzuti}{2011}]{deriglazov2011reparametrization}
\begin{barticle}
\bauthor{\bsnm{Deriglazov}, \binits{A.}},
\bauthor{\bsnm{Rizzuti}, \binits{B.}}:
\batitle{Reparametrization-invariant formulation of classical mechanics and the
  schr{\"o}dinger equation}.
\bjtitle{American Journal of Physics}
\bvolume{79}(\bissue{8}),
\bfpage{882}--\blpage{885}
(\byear{2011})
\end{barticle}
\endbibitem

\bibitem[\protect\citeauthoryear{Kronz and Lupher}{2005}]{Kronz2005}
\begin{barticle}
\bauthor{\bsnm{Kronz}, \binits{F.M.}},
\bauthor{\bsnm{Lupher}, \binits{T.A.}}:
\batitle{Unitarily inequivalent representations in algebraic quantum theory}.
\bjtitle{International Journal of Theoretical Physics}
\bvolume{44}(\bissue{8}),
\bfpage{1239}--\blpage{1258}
(\byear{2005})
\doiurl{10.1007/s10773-005-4683-0}
\end{barticle}
\endbibitem

\bibitem[\protect\citeauthoryear{Deriglazov and
  Evdokimov}{2000}]{deriglazov2000local}
\begin{barticle}
\bauthor{\bsnm{Deriglazov}, \binits{A.A.}},
\bauthor{\bsnm{Evdokimov}, \binits{K.E.}}:
\batitle{Local symmetries and the noether identities in the hamiltonian
  framework}.
\bjtitle{International Journal of Modern Physics A}
\bvolume{15}(\bissue{25}),
\bfpage{4045}--\blpage{4067}
(\byear{2000})
\end{barticle}
\endbibitem

\bibitem[\protect\citeauthoryear{Deriglazov}{2009}]{deriglazov2009improved}
\begin{barticle}
\bauthor{\bsnm{Deriglazov}, \binits{A.A.}}:
\batitle{Improved extended hamiltonian and search for local symmetries}.
\bjtitle{Journal of mathematical physics}
\bvolume{50}(\bissue{1}),
\bfpage{012907}
(\byear{2009})
\end{barticle}
\endbibitem

\bibitem[\protect\citeauthoryear{Deriglazov}{1995}]{deriglazov1995notes}
\begin{botherref}
\oauthor{\bsnm{Deriglazov}, \binits{A.}}:
Notes on lagrangean and hamiltonian symmetries.
arXiv preprint hep-th/9412244
(1995)
\end{botherref}
\endbibitem

\bibitem[\protect\citeauthoryear{DeWitt}{1967}]{dewitt1967quantum}
\begin{barticle}
\bauthor{\bsnm{DeWitt}, \binits{B.S.}}:
\batitle{Quantum theory of gravity. ii. the manifestly covariant theory}.
\bjtitle{Physical Review}
\bvolume{162}(\bissue{5}),
\bfpage{1195}
(\byear{1967})
\end{barticle}
\endbibitem

\bibitem[\protect\citeauthoryear{Dirac}{1950}]{dirac1950generalized}
\begin{barticle}
\bauthor{\bsnm{Dirac}, \binits{P.A.M.}}:
\batitle{Generalized hamiltonian dynamics}.
\bjtitle{Canadian journal of mathematics}
\bvolume{2},
\bfpage{129}--\blpage{148}
(\byear{1950})
\end{barticle}
\endbibitem

\bibitem[\protect\citeauthoryear{Dirac}{1988}]{dirac1988lectures}
\begin{barticle}
\bauthor{\bsnm{Dirac}, \binits{P.}}:
\batitle{Lectures on quantum mechanics (yeshiva univ., new york, 1964); ld
  faddeev and r. jackiw}.
\bjtitle{Phys. Rev. Lett}
\bvolume{60},
\bfpage{1692}
(\byear{1988})
\end{barticle}
\endbibitem

\bibitem[\protect\citeauthoryear{Page and Wootters}{1983}]{page1983evolution}
\begin{barticle}
\bauthor{\bsnm{Page}, \binits{D.N.}},
\bauthor{\bsnm{Wootters}, \binits{W.K.}}:
\batitle{Evolution without evolution: Dynamics described by stationary
  observables}.
\bjtitle{Physical Review D}
\bvolume{27}(\bissue{12}),
\bfpage{2885}
(\byear{1983})
\end{barticle}
\endbibitem

\bibitem[\protect\citeauthoryear{Wootters}{1984}]{wootters1984time}
\begin{barticle}
\bauthor{\bsnm{Wootters}, \binits{W.K.}}:
\batitle{“time” replaced by quantum correlations}.
\bjtitle{International journal of theoretical physics}
\bvolume{23}(\bissue{8}),
\bfpage{701}--\blpage{711}
(\byear{1984})
\end{barticle}
\endbibitem

\bibitem[\protect\citeauthoryear{Smith and Ahmadi}{2020}]{Smith:2019imm}
\begin{barticle}
\bauthor{\bsnm{Smith}, \binits{A.R.H.}},
\bauthor{\bsnm{Ahmadi}, \binits{M.}}:
\batitle{{Quantum clocks observe classical and quantum time dilation}}.
\bjtitle{Nature Commun.}
\bvolume{11}(\bissue{1}),
\bfpage{5360}
(\byear{2020})
\doiurl{10.1038/s41467-020-18264-4}
{\href{https://arxiv.org/abs/1904.12390}{{arXiv:1904.12390}}}
{[quant-ph]}
\end{barticle}
\endbibitem

\bibitem[\protect\citeauthoryear{Giacomini}{2021}]{Giacomini_2021}
\begin{barticle}
\bauthor{\bsnm{Giacomini}, \binits{F.}}:
\batitle{Spacetime quantum reference frames and superpositions of proper
  times}.
\bjtitle{Quantum}
\bvolume{5},
\bfpage{508}
(\byear{2021})
\doiurl{10.22331/q-2021-07-22-508}
\end{barticle}
\endbibitem

\bibitem[\protect\citeauthoryear{Rijavec}{2023}]{Rijavec_2023}
\begin{botherref}
\oauthor{\bsnm{Rijavec}, \binits{S.}}:
Robustness of the page-wootters construction across different pictures, states
  of the universe, and system-clock interactions.
Physical Review D
\textbf{108}(6)
(2023)
\doiurl{10.1103/physrevd.108.063507}
\end{botherref}
\endbibitem

\bibitem[\protect\citeauthoryear{Paczos et~al.}{2024}]{Paczos_2024}
\begin{barticle}
\bauthor{\bsnm{Paczos}, \binits{J.}},
\bauthor{\bsnm{Dębski}, \binits{K.}},
\bauthor{\bsnm{Grochowski}, \binits{P.T.}},
\bauthor{\bsnm{Smith}, \binits{A.R.H.}},
\bauthor{\bsnm{Dragan}, \binits{A.}}:
\batitle{Quantum time dilation in a gravitational field}.
\bjtitle{Quantum}
\bvolume{8},
\bfpage{1338}
(\byear{2024})
\doiurl{10.22331/q-2024-05-07-1338}
\end{barticle}
\endbibitem

\bibitem[\protect\citeauthoryear{Page}{1989}]{page1989itp}
\begin{botherref}
\oauthor{\bsnm{Page}, \binits{D.}}:
Itp preprint nsf-itp-89-18.
Time as an inaccessible observable
(1989)
\end{botherref}
\endbibitem

\bibitem[\protect\citeauthoryear{Giovannetti
  et~al.}{2015}]{giovannetti2015quantum}
\begin{barticle}
\bauthor{\bsnm{Giovannetti}, \binits{V.}},
\bauthor{\bsnm{Lloyd}, \binits{S.}},
\bauthor{\bsnm{Maccone}, \binits{L.}}:
\batitle{Quantum time}.
\bjtitle{Physical Review D}
\bvolume{92}(\bissue{4}),
\bfpage{045033}
(\byear{2015})
\end{barticle}
\endbibitem

\bibitem[\protect\citeauthoryear{Leon and Maccone}{2017}]{Leon_2017}
\begin{barticle}
\bauthor{\bsnm{Leon}, \binits{J.}},
\bauthor{\bsnm{Maccone}, \binits{L.}}:
\batitle{The pauli objection}.
\bjtitle{Foundations of Physics}
\bvolume{47}(\bissue{12}),
\bfpage{1597}--\blpage{1608}
(\byear{2017})
\doiurl{10.1007/s10701-017-0115-2}
\end{barticle}
\endbibitem

\bibitem[\protect\citeauthoryear{Busch et~al.}{1994}]{BUSCH1994357}
\begin{barticle}
\bauthor{\bsnm{Busch}, \binits{P.}},
\bauthor{\bsnm{Grabowski}, \binits{M.}},
\bauthor{\bsnm{Lahti}, \binits{P.J.}}:
\batitle{Time observables in quantum theory}.
\bjtitle{Physics Letters A}
\bvolume{191}(\bissue{5}),
\bfpage{357}--\blpage{361}
(\byear{1994})
\doiurl{10.1016/0375-9601(94)90785-4}
\end{barticle}
\endbibitem

\bibitem[\protect\citeauthoryear{Loveridge and Miyadera}{2019}]{Loveridge_2019}
\begin{barticle}
\bauthor{\bsnm{Loveridge}, \binits{L.}},
\bauthor{\bsnm{Miyadera}, \binits{T.}}:
\batitle{Relative quantum time}.
\bjtitle{Foundations of Physics}
\bvolume{49}(\bissue{6}),
\bfpage{549}--\blpage{560}
(\byear{2019})
\doiurl{10.1007/s10701-019-00268-w}
\end{barticle}
\endbibitem

\bibitem[\protect\citeauthoryear{Singh and Carroll}{2018}]{Singh:2018qzk}
\begin{botherref}
\oauthor{\bsnm{Singh}, \binits{A.}},
\oauthor{\bsnm{Carroll}, \binits{S.M.}}:
{Modeling Position and Momentum in Finite-Dimensional Hilbert Spaces via
  Generalized Clifford Algebra}
(2018)
{\href{https://arxiv.org/abs/1806.10134}{{arXiv:1806.10134}}}
{[quant-ph]}
\end{botherref}
\endbibitem

\bibitem[\protect\citeauthoryear{H\"ohn et~al.}{2021}]{PhysRevD.104.066001}
\begin{barticle}
\bauthor{\bsnm{H\"ohn}, \binits{P.A.}},
\bauthor{\bsnm{Smith}, \binits{A.R.H.}},
\bauthor{\bsnm{Lock}, \binits{M.P.E.}}:
\batitle{Trinity of relational quantum dynamics}.
\bjtitle{Phys. Rev. D}
\bvolume{104},
\bfpage{066001}
(\byear{2021})
\doiurl{10.1103/PhysRevD.104.066001}
\end{barticle}
\endbibitem

\bibitem[\protect\citeauthoryear{Favalli and
  Smerzi}{2020}]{Favalli2020timeobservablesin}
\begin{barticle}
\bauthor{\bsnm{Favalli}, \binits{T.}},
\bauthor{\bsnm{Smerzi}, \binits{A.}}:
\batitle{Time {O}bservables in a {T}imeless {U}niverse}.
\bjtitle{{Quantum}}
\bvolume{4},
\bfpage{354}
(\byear{2020})
\doiurl{10.22331/q-2020-10-29-354}
\end{barticle}
\endbibitem

\bibitem[\protect\citeauthoryear{Ruiz et~al.}{2017}]{CastroRuiz2017}
\begin{barticle}
\bauthor{\bsnm{Ruiz}, \binits{E.C.}},
\bauthor{\bsnm{Giacomini}, \binits{F.}},
\bauthor{\bsnm{Brukner}}:
\batitle{Entanglement of quantum clocks through gravity}.
\bjtitle{Proceedings of the National Academy of Sciences}
\bvolume{114}(\bissue{12}),
\bfpage{2303}--\blpage{2309}
(\byear{2017})
\doiurl{10.1073/pnas.1616427114}
{\href{https://arxiv.org/abs/https://www.pnas.org/doi/pdf/10.1073/pnas.1616427114}{{https://www.pnas.org/doi/pdf/10.1073/pnas.1616427114}}}
\end{barticle}
\endbibitem

\bibitem[\protect\citeauthoryear{Castro-Ruiz et~al.}{2020}]{Castro_Ruiz_2020}
\begin{botherref}
\oauthor{\bsnm{Castro-Ruiz}, \binits{E.}},
\oauthor{\bsnm{Giacomini}, \binits{F.}},
\oauthor{\bsnm{Belenchia}, \binits{A.}},
\oauthor{\bsnm{Caslav}, \binits{B.}}:
Quantum clocks and the temporal localisability of events in the presence of
  gravitating quantum systems.
Nature Communications
\textbf{11}(1)
(2020)
\doiurl{10.1038/s41467-020-16013-1}
\end{botherref}
\endbibitem

\bibitem[\protect\citeauthoryear{Smith and Ahmadi}{2019}]{smith2019quantizing}
\begin{barticle}
\bauthor{\bsnm{Smith}, \binits{A.R.}},
\bauthor{\bsnm{Ahmadi}, \binits{M.}}:
\batitle{Quantizing time: Interacting clocks and systems}.
\bjtitle{Quantum}
\bvolume{3},
\bfpage{160}
(\byear{2019})
\end{barticle}
\endbibitem

\bibitem[\protect\citeauthoryear{Marletto and Vedral}{2022}]{Marletto_2022}
\begin{botherref}
\oauthor{\bsnm{Marletto}, \binits{C.}},
\oauthor{\bsnm{Vedral}, \binits{V.}}:
The quantum totalitarian property and exact symmetries.
AVS Quantum Science
\textbf{4}(1)
(2022)
\doiurl{10.1116/5.0077192}
\end{botherref}
\endbibitem

\bibitem[\protect\citeauthoryear{Singh}{2022}]{Singh:2020kdu}
\begin{barticle}
\bauthor{\bsnm{Singh}, \binits{A.}}:
\batitle{{Quantum space, quantum time, and relativistic quantum mechanics}}.
\bjtitle{Quant. Stud. Math. Found.}
\bvolume{9}(\bissue{1}),
\bfpage{35}--\blpage{53}
(\byear{2022})
\doiurl{10.1007/s40509-021-00255-9}
{\href{https://arxiv.org/abs/2004.09139}{{arXiv:2004.09139}}}
{[quant-ph]}
\end{barticle}
\endbibitem

\bibitem[\protect\citeauthoryear{Maggiore}{1993}]{Maggiore:1993rv}
\begin{barticle}
\bauthor{\bsnm{Maggiore}, \binits{M.}}:
\batitle{{A Generalized uncertainty principle in quantum gravity}}.
\bjtitle{Phys. Lett. B}
\bvolume{304},
\bfpage{65}--\blpage{69}
(\byear{1993})
\doiurl{10.1016/0370-2693(93)91401-8}
{\href{https://arxiv.org/abs/hep-th/9301067}{{arXiv:hep-th/9301067}}}
\end{barticle}
\endbibitem

\bibitem[\protect\citeauthoryear{Ali et~al.}{2009}]{Ali:2009zq}
\begin{barticle}
\bauthor{\bsnm{Ali}, \binits{A.F.}},
\bauthor{\bsnm{Das}, \binits{S.}},
\bauthor{\bsnm{Vagenas}, \binits{E.C.}}:
\batitle{{Discreteness of Space from the Generalized Uncertainty Principle}}.
\bjtitle{Phys. Lett. B}
\bvolume{678},
\bfpage{497}--\blpage{499}
(\byear{2009})
\doiurl{10.1016/j.physletb.2009.06.061}
{\href{https://arxiv.org/abs/0906.5396}{{arXiv:0906.5396}}}
{[hep-th]}
\end{barticle}
\endbibitem

\bibitem[\protect\citeauthoryear{Ali et~al.}{2011}]{Ali:2011fa}
\begin{barticle}
\bauthor{\bsnm{Ali}, \binits{A.F.}},
\bauthor{\bsnm{Das}, \binits{S.}},
\bauthor{\bsnm{Vagenas}, \binits{E.C.}}:
\batitle{{A proposal for testing Quantum Gravity in the lab}}.
\bjtitle{Phys. Rev. D}
\bvolume{84},
\bfpage{044013}
(\byear{2011})
\doiurl{10.1103/PhysRevD.84.044013}
{\href{https://arxiv.org/abs/1107.3164}{{arXiv:1107.3164}}}
{[hep-th]}
\end{barticle}
\endbibitem

\bibitem[\protect\citeauthoryear{Tino}{2021}]{Tino_2021}
\begin{barticle}
\bauthor{\bsnm{Tino}, \binits{G.M.}}:
\batitle{Testing gravity with cold atom interferometry: results and prospects}.
\bjtitle{Quantum Science and Technology}
\bvolume{6}(\bissue{2}),
\bfpage{024014}
(\byear{2021})
\doiurl{10.1088/2058-9565/abd83e}
\end{barticle}
\endbibitem

\end{thebibliography}

\end{document}